\documentclass[onecolumn]{revtex4}
\usepackage{amsmath,amssymb,graphics,epsfig,subfigure}
\usepackage{color}
\usepackage[colorlinks,linkcolor=red,anchorcolor=red,citecolor=green]{hyperref}
\usepackage{setspace}
\usepackage{booktabs}
\usepackage{float}
\usepackage{appendix}
\usepackage{makecell}
\usepackage{multirow}
\begin{document}

\thispagestyle{empty}

\begin{center}

\title{Thermodynamic bounce effect in quantum BTZ black hole}

\author{Zhen-Ming Xu\footnote{E-mail: zmxu@nwu.edu.cn}, Pan-Pan Zhang, Bin Wu, Xing Zhang
        \vspace{6pt}\\}

\affiliation{$^{1}$School of Physics, Northwest University, Xi'an 710127, China\\
$^{2}$Shaanxi Key Laboratory for Theoretical Physics Frontiers, Xi'an 710127, China\\
$^{3}$Peng Huanwu Center for Fundamental Theory, Xi'an 710127, China}

\begin{abstract}
A novel thermodynamic phenomenon has been observed in the quantum Ba\~{n}ados-Teitelboim-Zanelli (qBTZ) black hole, utilizing generalized free energy and Kramer escape rate. This phenomenon also reveals the unique property of the quantum black hole. The stochastic thermal motion of various thermodynamic states within the black hole system induces phase transitions, under the influence of generalized free energy which obtained by extending Maxwell's construction. Through the analysis of Kramer escape rate, it is discovered that the qBTZ black hole thermodynamic system exhibits a bounce effect. It originates from the non-monotonicity of entropy
in black hole thermodynamic systems. Furthermore, the overall thermodynamic picture of the qBTZ black hole has been obtained under different quantum backreactions.
\end{abstract}

\maketitle
\end{center}

\section{Introduction}
The proposal that black holes have properties of temperature (related to the surface gravity of black holes) and entropy (proportional to the event horizon area of black holes) makes the black hole an effective window for quantum gravity~\cite{Hawking1974,Bekenstein1973,Hawking1976}. With the discovery of the gravitational wave and the rapid development of quantum computation, on the one hand, people have used gravitational wave data to test the Hawking's black-hole area theorem~\cite{Isi2021}, and on the other hand, people have used superconducting quantum computation platform to simulate the Hawking radiation and dynamics of the curved spacetime~\cite{Shi2023}. These works indirectly support the development of black hole thermodynamics.

Looking only at the black hole thermodynamics, there have been significant theoretical advancements recently, which have expanded the research content of black hole thermodynamics and deepened our understanding of the underlying structure of black hole thermodynamics. From the initial examination of the quantitative relationship satisfied by some thermodynamic quantities of black holes~\cite{Bardeen1973} to the proposal of extended phase space or black hole chemistry~\cite{Kastor2009,Dolan2011,Kubiznak2012,Kubiznak2017}, black hole thermodynamic systems exhibit rich thermodynamic critical behaviors, greatly expanding the research scope of black hole physics. The development of holographic interpretation of black hole thermodynamics~\cite{Visser2022,Cong2022,Gao2022,Frassino2023} provides a great theoretical reference for viewing the black hole system as a {\em more real} thermodynamic system.

The hypothesis of black hole molecule~\cite{Wei2015} has largely guided the phenomenological description of the microscopic mechanisms~\cite{Miao2018,Wei2019a,Wei2019b,Xu2020,Ghosh2020} and the dynamic process~\cite{Li2020,Wei2021,Cai2021} behind the rich thermodynamic phenomena of black holes. At the same time, the study of the thermodynamic topology of black holes~\cite{Wei2022a,Wei2022b}, description of coexistence-zone~\cite{Wei2024} and the application of extended Iyer-Wald formalism~\cite{Caceres2017,Couch2017,Xiao2024,Hajian2024} in black hole thermodynamics have increased the macroscopic understanding of the thermodynamic phenomena of black holes.

The macroscopic thermal phenomena exhibited by black hole thermodynamic systems are our main means of analyzing the physical properties of a black hole. Especially the discovery of thermodynamic phase transitions in black holes has greatly promoted people's understanding of black hole physics. In this study, we find a thermodynamic phase transition behavior that has never been reported in literature before, which we refer to as thermodynamic bounce effect. In short, with the continuous increasing of thermodynamic quantity, the first-order phase transition of the system exhibits the characteristic of ``existence-disappear-existence'', and the phase transition rate starts to decrease from a certain value to zero, and then starts to increase from zero to the maximum value before gradually decreasing again to zero.

The discovery of this interesting thermodynamic phenomenon is due to the recent proposal of the qBTZ black hole~\cite{Emparan2020,Panella2024}, which is an exact solution to specific semiclassical gravitational equations due to quantum conformal matter in the context of holographic braneworld models. In this sence, it may be understood as a quantum black hole with all orders in quantum backreaction. It is precisely because of this correction from quantum backreaction on the brane that the black hole thermodynamic system exhibits thus anomalous behavior. The qBTZ black hole has attracted people's attention due to its unique characteristics. Reentrant phase transition~\cite{Frassino2024a}, quantum inequalities~\cite{Frassino2024b}, specific heats in extended thermodynamics~\cite{Johnson2023}, thermodynamical topology~\cite{Wu2024} and the thermodynamic geometry~\cite{Mansoori2024} have been investigated in detail. Here we have observed another thermodynamic phenomenon for this black hole, i.e. thermodynamic bounce effect. Based on the existing analysis in the literature, we have provided an overall thermodynamic picture of the thermodynamic system of the qBTZ black hole, deepening our understanding of quantum black holes.

Thus, a new thermodynamic analysis method we adopt is to view the phase transition of black holes as caused by Brownian motion, and use the Fokker-Planck equation in stochastic motion for analysis, thereby obtaining more details in the thermal motion of the system. Next, we will elaborate on our proposed analysis scheme.

\section{Transition Process}
The Brownian motion of particles in multiple potential wells will undergo transitions between potential wells under low noise conditions. What we are more concerned about is the transition process between two steady states. The evolution problem of the system from steady state under the stochastic force, satisfying the Fokker-Planck equation~\cite{Risken1988},
\begin{equation}\label{fpeq}
\frac{\partial P(z,t)}{\partial t}=L_{\text{FP}}P(z,t), \qquad L_{\text{FP}}=D\frac{\partial}{\partial z}e^{-U(z)/D}\frac{\partial}{\partial z}e^{U(z)/D},
\end{equation}
where the probability distribution $P(z,t)$, the constant diffusion coefficient $D$ and the potential $U(z)$. For simplicity, in this study, we currently consider that when a particle crosses a potential barrier (at $z_2$) through a transition in a potential well (at $z_1$), its probability distribution will become zero, i.e., $P(A,t)=0$ for $A>z_2$, which indicates that we do not care about the specific distribution of the particle after crossing the potential barrier, nor do we consider the backflow problem during the transition. Meanwhile, we also think that the transition process takes place in a system where local equilibrium has already been formed within the region of $z<A$ for $A>z_2$.

Based on these considerations, we can ultimately obtain the total probability distribution $M(t)$ that the particle is in the interval $(-\infty, A)$ at time $t$ as
\begin{equation}
M(t)=e^{-vt}, \qquad \frac{1}{v}=\frac{1}{D}\int_{-\infty}^{A}e^{-U(z)/D}dz\int_{z_1}^{A}e^{U(z)/D}dz.
\end{equation}
Under the weak noise condition, that is the small value of constant diffusion $D$, the main contribution of the above two integrals comes from the contributions within the corresponding extremum point neighborhood. The Taylor expansions of the potential function $U(z)$ near two extreme points are
\begin{align}
U(z)&=U(z_1)+\frac12 U''(z_1)(z-z_1)^2+\cdots,\\
U(z)&=U(z_2)-\frac12 |U''(z_2)|(z-z_2)^2+\cdots.
\end{align}
Hence we can obtain the expression for the transition rate~\cite{Risken1988,Xu2023,Liu2023,Du2023,Wang2024,Sadeghi2024}
\begin{equation}\label{rate}
	v=\frac{1}{2\pi}\sqrt{|U''(z_1)U''(z_2)|}e^{-\Delta U/D}, \qquad \Delta U=U(z_2)-U(z_1),
\end{equation}
and the above equation is also the Kramers escape rate.

Naturally, how to construct potential becomes a key issue, especially for the thermodynamic systems of black holes we are examining. Fortunately, in our previous work~\cite{Xu2024,Xu2021a,Wang2023}, we constructed the generalized free energy of a black hole thermodynamic system, placing the different states in the system at the extremum of the potential. Our starting point for constructing generalized free energy is the Maxwell's equal area law in classical thermodynamics, which is extended to obtain the potential~\cite{Xu2024}
\begin{equation}\label{poten}
	U(z)=\int(T_h(z)-T)\text{d}S(z),
\end{equation}
here, $T_h(z)$ is the Hawking temperature of the black hole thermodynamic system, $S(z)$ is the Bekenstein-Hawking entropy of the black hole, both of which are functions of the radius $z$ of the black hole event horizon, and $T$ is a free parameter that can be taken as any positive value in any way. In the study~\cite{Xu2021a}, the parameter $T$ is just the temperature of a canonical ensemble composed of a large number of states, including the  on-shell black hole state and the off-shell unknown states.

\section{Thermodynamic bounce effect}
We start our analysis with the quantum-corrected black holes to all orders of backreaction in semiclassical gravity. These black holes are exist with the framework of an AdS$_3$ end-of-the-world brane intersecting an appropriate AdS$_4$ $C-$metric black hole horizon, and can be regarded as qBTZ black holes~\cite{Emparan2020,Panella2024}. By braneworld holography, we can obtained some thermodynamics of this black hole analytically from the AdS$_4$ bulk black hole thermodynamics. The mass, entropy and the temperature of the qBTZ black hole are~\cite{Frassino2024a,Frassino2024b,Johnson2023}
\begin{equation}\label{mass}
  M(z,\nu)=\frac{\sqrt{1+\nu^2}}{2G_3}\frac{z^2(1-\nu z^3)(1+\nu z)}{(1+3z^2+2\nu z^3)^2},
\end{equation}
\begin{equation}\label{entropy}
  S(z,\nu)=\frac{\pi l_3\sqrt{1+\nu^2}}{G_3}\frac{z}{1+3z^2+2\nu z^3},
\end{equation}
\begin{equation}\label{temp}
  T_h(z,\nu)=\frac{1}{2\pi l_3}\frac{z(2+3\nu z+\nu z^3)}{1+3z^2+2\nu z^3},
\end{equation}
where $\nu=l/l_3$ and $z=l_3/(r_h x)$ for the UV cut-off length scale $l$, the geometric parameter $x$ of the bulk $C-$metric, the horizon radius $r_h$ and AdS$_3$ length $l_3$ with related to the brane cosmological constant, and both $\nu$ and $z$ have range $[0, \infty)$.  From the perspective that the qBTZ black hole can be regarded as a solution to the induced semi-classical theory on the brane at all orders in quantum backreaction,  the UV cut-off length scale $l$ controls the strength of backreaction. With fixed the AdS$_3$ length $l_3$, we can clearly see that $\nu \ll 1$ means  small backreaction, $\nu \gg 1$ means large backreaction and $\nu =0$ means  vanishing backreaction.

For the generalized free energy or potential of the qBTZ black hole, it takes the form with the help of the Eqs.~(\ref{poten}),~(\ref{entropy}) and~(\ref{temp}),
\begin{equation}\label{gfree}
	u=\frac{\sqrt{1+\nu^2} \left[\left(z^2+1\right) \left(8 \nu z^3+9 z^2+1\right)-8t z \left(2 \nu z^3+3 z^2+1\right)\right]}{8\left(2 \nu z^3+3 z^2+1\right)^2},
\end{equation}
where $U(z,\nu)=\int (T_h(z,\nu)-T)\text{d}S(z,\nu)$ with the rescaling dimensionless quantities $u=G_3 U(z,\nu)$ and $t=\pi l_3 T$.

By utilizing the generalized free energy or potential Eq.~(\ref{gfree}) and transition rate or Kramers escape rate Eq.~(\ref{rate}), we can obtain the process information of the qBTZ black hole phase transition, which can be simply divided into four cases with different backreaction parameter $\nu$.

\begin{figure}[htbp]
	\centering
	\subfigure[$\nu=0.5$]{\includegraphics[width=15cm]{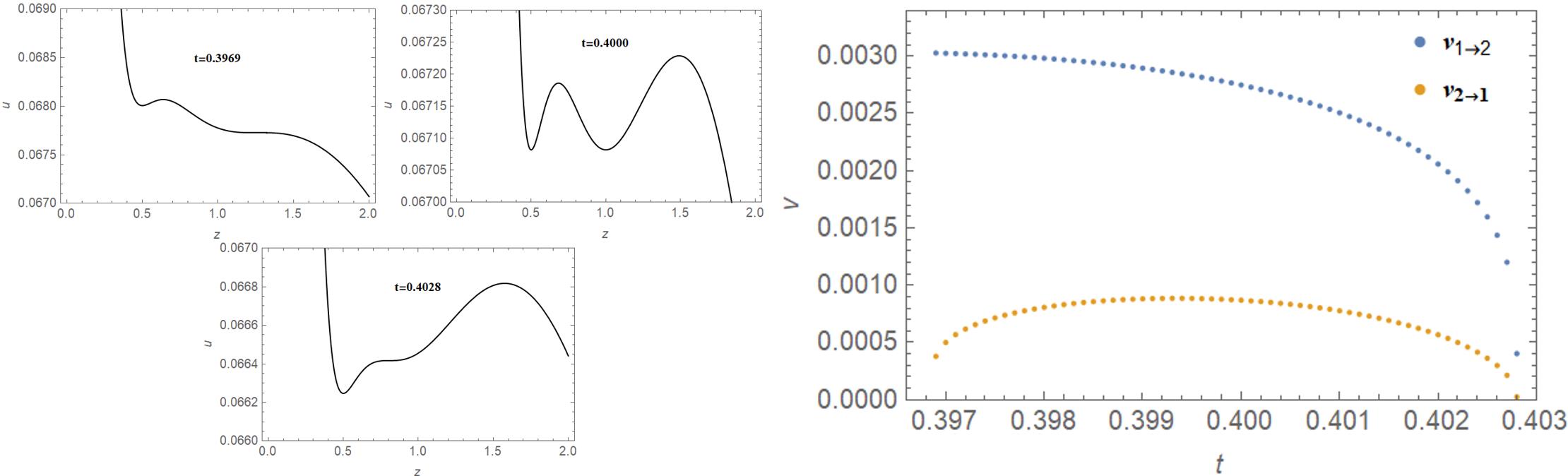}}
	\subfigure[$\nu=1.0$]{\includegraphics[width=15cm]{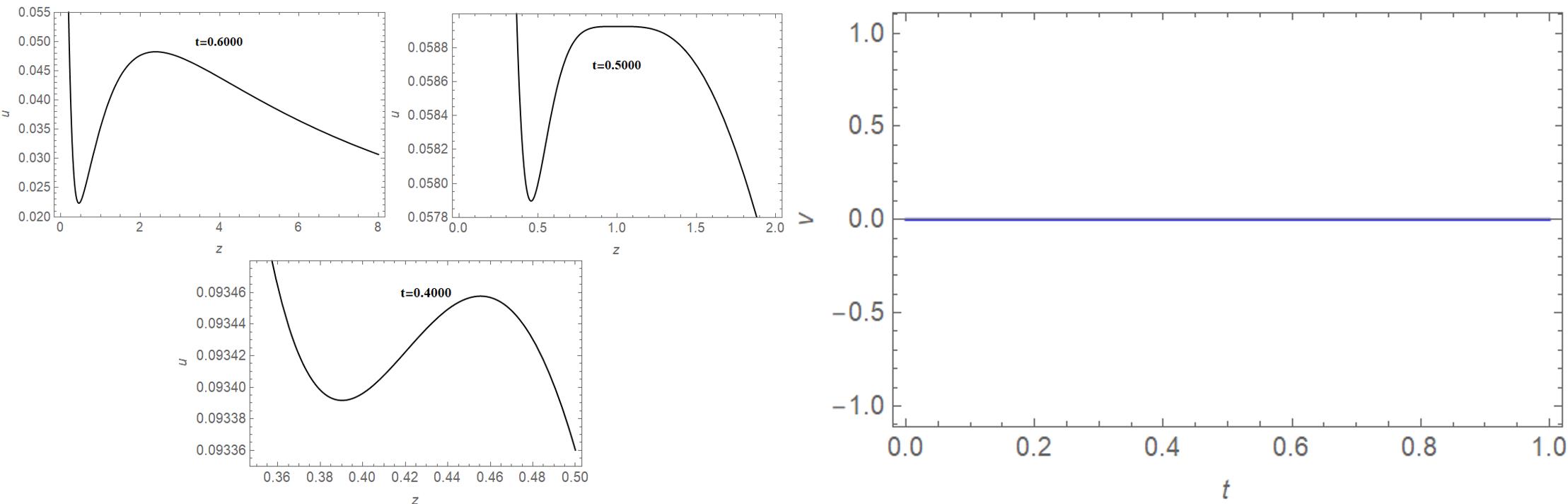}}
	\subfigure[$\nu=5.0$]{\includegraphics[width=15cm]{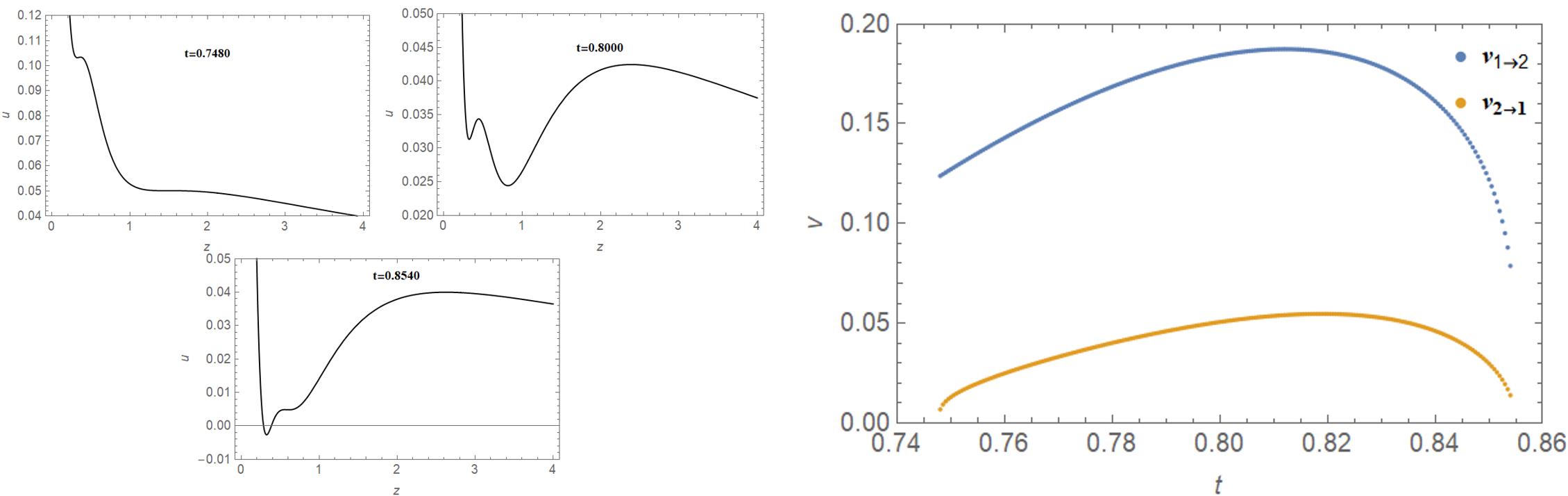}}
    \subfigure[$\nu=7.5$]{\includegraphics[width=15cm]{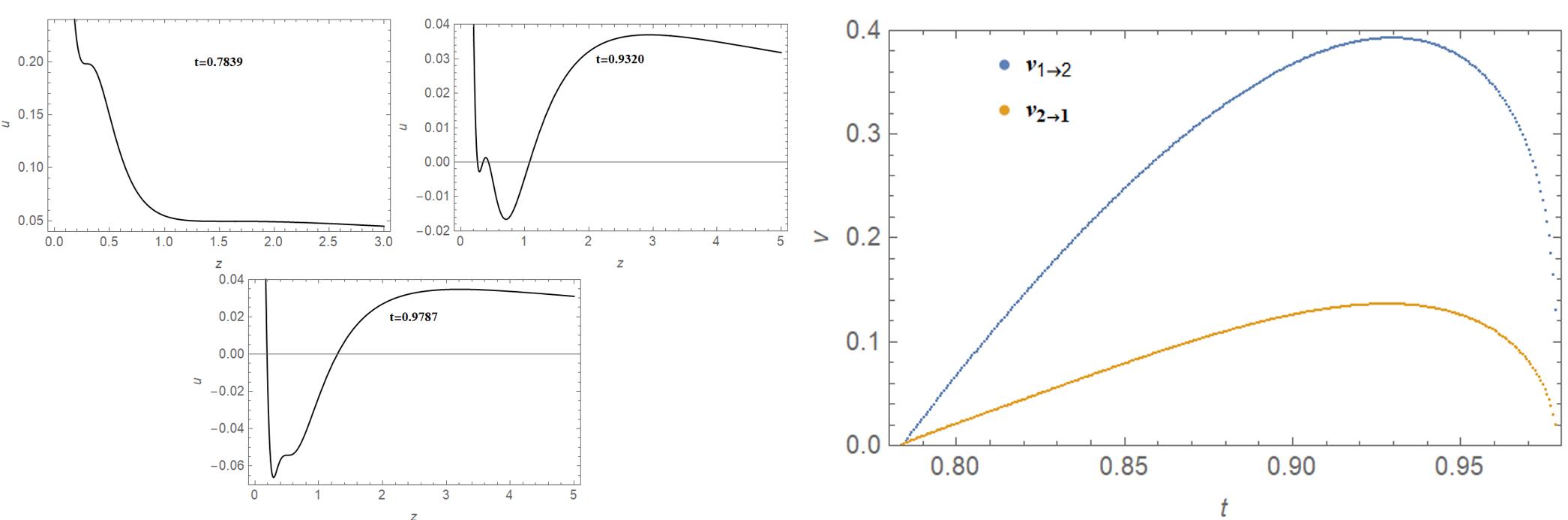}}
	\caption{{\bf Left half part: }The behavior of generalized free energy with respect to the event horizon radius of the black hole within a certain interval of the temperature. {\bf Right half part: }The behavior of the phase transition rate with respect to the temperature, where $v_{1\rightarrow 2}$ represents the transition of a particle or state from the leftmost potential well of generalized free energy to the rightmost one, while $v_{2\rightarrow 1}$ represents the opposite process.}
	\label{fig1}
\end{figure}

\textbf{(1) Case $\nu<1$: first-order phase transition}

We have plotted the behavior of generalized free energy and transition rate in diagram (a) for Fig.~\ref{fig1}. For the generalized free energy, it can be seen that within a certain range of the temperature, it has two global minimum points, and these two minimum points will transform into each other, indicating a first-order phase transition in the system. The corresponding phase transition rate shows a gradually decreasing trend as the temperature increases.

\textbf{(2) Case $\nu=1$: no phase transition}

In diagram (b) for Fig.~\ref{fig1}, we can clearly see that when $\nu=1$, the position where the extreme point originally existed in Case (1) becomes a platform, and the behavior of the entire generalized free energy only has a global minimum, indicating that there will be no phase transition in the system, and the corresponding phase transition rate remains constant at zero.

\textbf{(3) Case $1<\nu \le \nu_c$: first-order phase transition}

In the case of the diagram (c) for Fig.~\ref{fig1}, at this point, the thermodynamic behavior of the qBTZ black hole returns to Case (1), which is the first-order phase transition. For the generalized free energy, it has two global minimum points, and two of them will transform into each other within a certain interval of the temperature. Correspondingly, the phase transition rate is non-zero and shows a trend of first increasing and then decreasing. Especially when $\nu=\nu_c$ shown in the diagram (d) for Fig.~\ref{fig1}, the phase transition rate exhibits a behavior of increasing from zero and then decreasing to zero, which opens up the thermodynamic bounce effect to be analyzed soon.

\textbf{(4) Case $\nu > \nu_c$: thermodynamic bounce effect}

\begin{figure}[htbp]
	\centering
	\includegraphics[width=18cm]{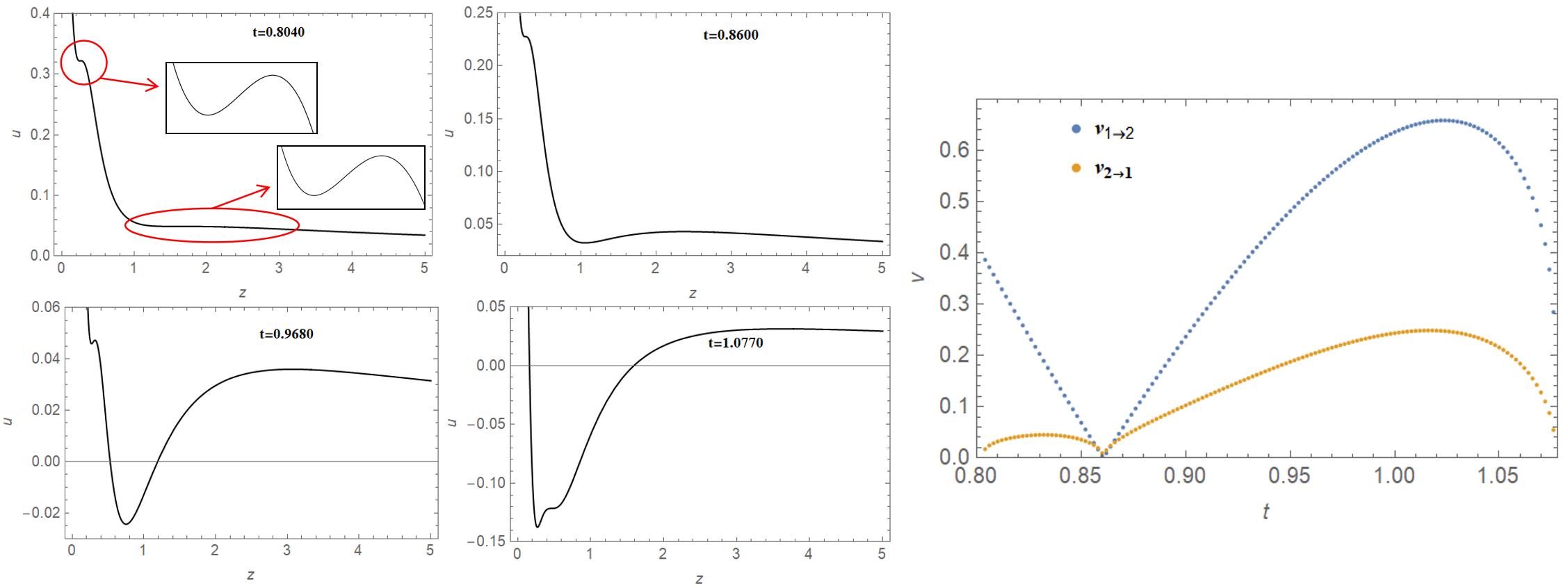}
	\caption{For $\nu=10.0 > \nu_c$, the specific process of thermodynamic bounce effect, including the behavior of generalized free energy and phase transition rate.}
	\label{fig2}.
\end{figure}

In this case, a unique phenomenon emerges in the thermodynamic phase transition of the qBTZ black hole, which we call thermodynamic bounce effect. Of course, this is for the space spanned by the phase transition rate and temperature, shown in Fig.~\ref{fig2}. For the generalized free energy, it has four extreme points. As the temperature increases, the first two extreme points can form a turning point, which is quite special. Once the turning point is generated, it immediately disappears and decomposes into two extreme points (local maximum and local minimum). This special behavior leads to a phenomenon where as the temperature increases, the system undergoes a first-order phase transition, which suddenly disappears at a certain temperature (like the $t=0.86$ in Fig.~\ref{fig2}) and as the temperature increases again, the first-order phase transition begins again. This result is reflected in the phase transition rate, showing a ``bounce'' phenomenon as shown in the right part of Fig.~\ref{fig2}. As the temperature continues to increase, the phase transition rate starts to decrease from a certain value to zero, and then starts to increase from zero to the maximum value before gradually decreasing again to zero.

The canonical system we are considering here is consistent with the one in~\cite{Frassino2024a}, which is simply fixed backreaction $\nu$, that is, a fixed pressure term in terms of the spirit of extended phase space. If we consider Gibbs free energy $G=M-TS$, we can't see this kind of bounce effect by plotting the relationship between Gibbs free energy and temperature. As shown in~\cite{Frassino2024a}, when $\nu>1$, the Gibbs free energy always has only one swallowtail behavior, indicating that a first-order phase transition will occur. Next, we will explain this situation in detail.

For the Gibbs free energy, according to the calculation in~\cite{Frassino2024a}, it is
      \begin{equation}
        G=M-TS=G_{\text{qBTZ}}-G_{\text{qTAdS}}=\frac{\sqrt{\nu^2+1}}{8G_3}-\frac{z^2 \sqrt{\nu^2+1} \left(\nu^2 z^4+2 \nu \left(z^3+z\right)+1\right)}{2G_3 \left(2 \nu z^3+3 z^2+1\right)^2}.
      \end{equation}
For the generalized free energy Eq.~(\ref{gfree}) we introduced, we make the free parameter $T$ equal to the Hawking temperature of the qBTZ black hole, i.e.,
  $$T=T_h=\frac{1}{2\pi l_3}\frac{z(2+3\nu z+\nu z^3)}{1+3z^2+2\nu z^3},$$ we obtain $U(z)|_{T=T_h}=G$. From this, we can see that our calculation scheme for free energy is equivalent to that in~\cite{Frassino2024a}. So why can't we see the thermodynamic bounce effect from the perspective of Gibbs free energy. This is because for qBTZ black hole, its thermodynamic entropy is a non-monotonic function with respect to the parameter $z$. It is due to the non-monotonicity that leads to the occurrence of bounce effect. Through the perspective of Gibbs free energy, this non-monotonicity has been smoothed out, so this novel thermodynamic effect cannot be manifested. Now let's take a closer look at the definition of generalized free energy, which can be rewritten as
    \begin{equation}
	U(z)=\int(T_h(z)-T)F(z)\text{d}z,
    \end{equation}
  where $\text{d}S(z)=F(z)\text{d}z$. The monotonicity or non-monotonicity of entropy is completely determined by the function $F(z)$. From the perspective of generalized free energy, we are concerned with the behavior of the extreme points of the generalized free energy, that is, they satisfy the condition $(T_h(z)-T)F(z)=0$. There are two situations here.
  \begin{itemize}
    \item $T_h(z)-T=0$, this situation corresponds to the viewpoint of Gibbs free energy. We have eliminated the influence of $F(z)$ in this plan.
    \item $F(z)=0$, it corresponds to the mathematical property of entropy that we need to consider.
  \end{itemize}
  For the generalized free energy scheme we currently use, we consider the effects of above two scenarios on the phase transition behavior of black hole thermodynamic systems simultaneously. The $T_h(z)-T=0$ means the intersection of the plot of black hole temperature $T_h(z)$ with respect to the parameter $z$ and different isotherms $T$. The $F(z)=0$ means the isotherm corresponding to the extreme point of entropy $S(z)$.

\begin{figure}[htbp]
	\centering
	\subfigure[$\nu=0.5$]{\includegraphics[width=8cm]{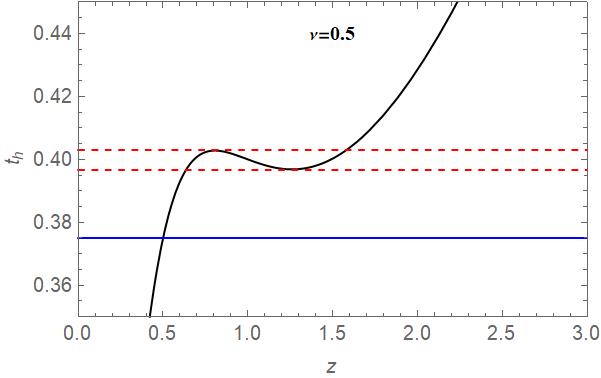}}
	\subfigure[$\nu=1.0$]{\includegraphics[width=8cm]{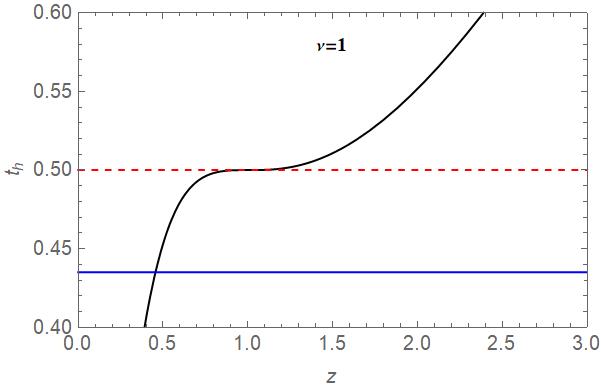}}
	\subfigure[$\nu=5.0$]{\includegraphics[width=8cm]{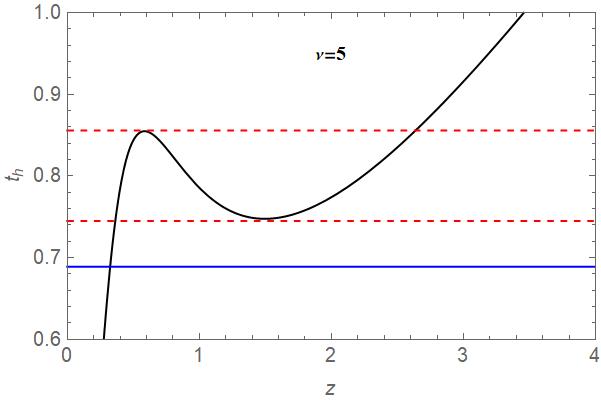}}
    \subfigure[$\nu=7.5$]{\includegraphics[width=8cm]{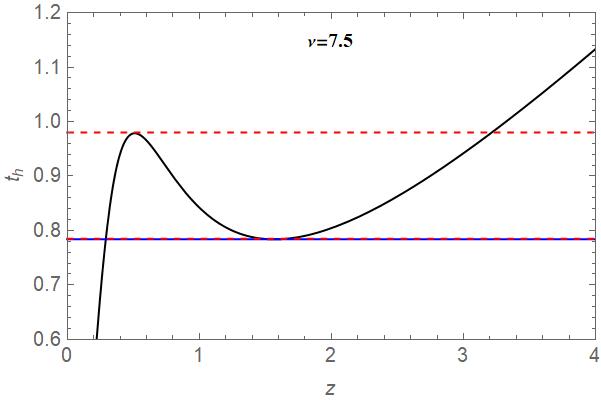}}
    \subfigure[$\nu=10$]{\includegraphics[width=8cm]{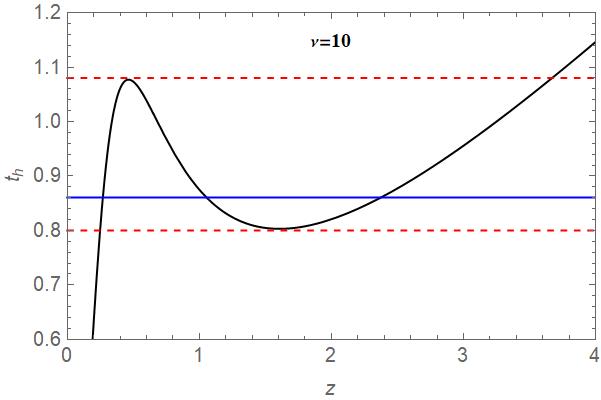}}
	\caption{The black solid line represents the behavior of black hole temperature $t_h$ with respect to $z$ for different $\nu$. Between the two red dashed lines, it corresponds to situation $T_h(z)-T=0$, where the phase transition occurs that we are concerned about. The blue solid line corresponds to the situation $F(z)=0$.}
\label{f1}
\end{figure}

  For qBTZ black hole, we have
  \begin{equation}
  t_h=\frac{z(2+3\nu z+\nu z^3)}{2(1+3z^2+2\nu z^3)},\quad  f(z)=-\frac{\sqrt{1 + \nu^2}(4 \nu z^3+3 z^2-1)}{(1 + 3 z^2 + 2 \nu z^3)^2},
  \end{equation}
  where $t_h=\pi l_3 T_h(z)$, $f(z)=G_3 F(z)/(\pi l_3)$. Here we have plotted the behavior of black hole temperature with respect to $z$ for different $\nu$. From the Fig.~\ref{f1}, we can see that as $\nu$ continues to increase, the non-monotonicity of entropy plays an important role, leading to the emergence of thermodynamic bounce effects.

\section{Discussion and Summary}\label{sec4}
\begin{figure}[htb]
\includegraphics[width=18cm]{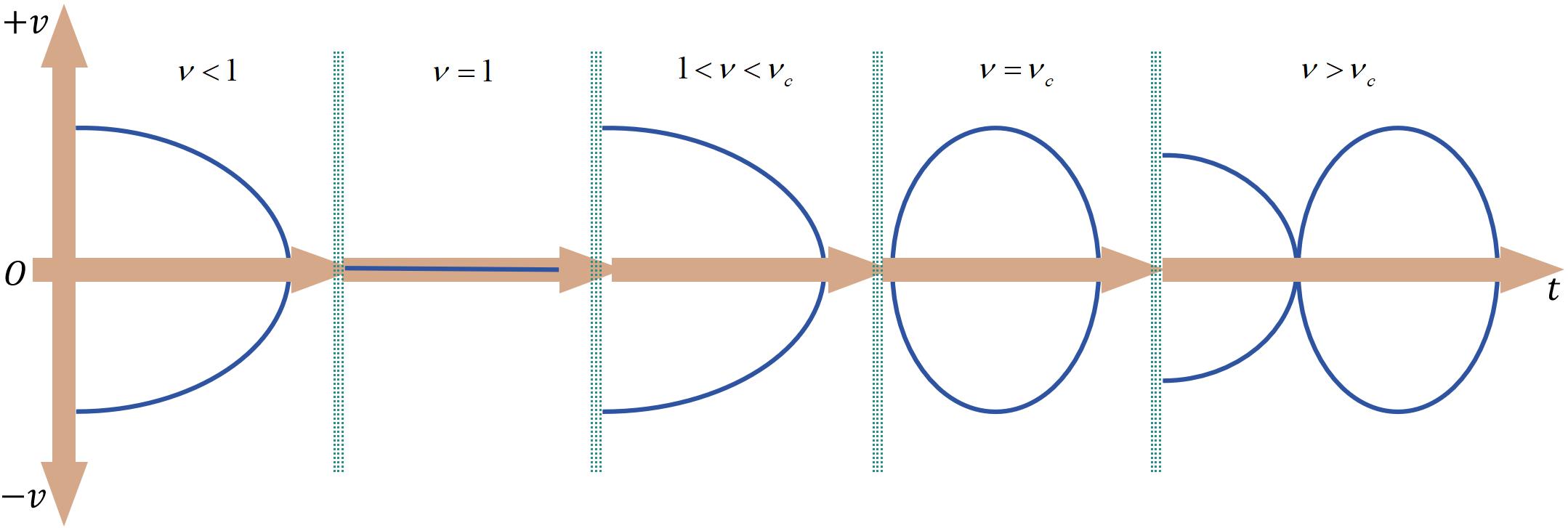}
	\caption{The overall schematic diagram of the thermodynamic process of the qBTZ black hole.}
	\label{fig3}
\end{figure}

In this study, we investigate the thermodynamic behavior of the qBTZ black hole. Although there have been relevant discussions in the literature and it has been pointed out that this black hole exhibits reentrant phase transitions~\cite{Frassino2024a}, as well as the specificity of $\nu=1$~\cite{Johnson2023}. In our research, we have discovered another interesting and unprecedented thermodynamic phenomenon called the thermodynamic bounce effect. To our knowledge, this is not observed in any black hole systems. The discovery of this special phenomenon implies the existence of another crucial value $\nu=\nu_c$ (like the $\nu_c=7.5$ in Fig.~\ref{fig1}) for the qBTZ black hole. Below $\nu_c$, the system will only exhibit a pure first-order phase transition, while above $\nu_c$, the phase transition behavior will be more unique, emerging with the characteristic of ``existence-disappear-existence''. Based on this, we can provide a comprehensive view of the thermodynamic characteristics of the quantum-corrected black holes with all orders of backreaction in semiclassical gravity, as shown in Fig.~\ref{fig3}. Here is a schematic diagram of the phase transition rate of the qBTZ black hole as a function of temperature, under different backreaction.

We have known that the three-dimensional qBTZ black hole is an exact solution to specific semiclassical gravitational equations due to quantum conformal matter, discovered through braneworld holography. The effect of the semiclassical backreaction is mainly reflected in the parameter $\nu$. A large $\nu$ indicates a large backreaction, while a small $\nu$ indicates a small backreaction. Focusing on the canonical ensemble with fixed $\nu$, the semiclassical backreaction has a significant impact on the thermodynamic behavior of the qBTZ black holes, which can be roughly divided into two different stages. The first stage is based on $\nu=1$ as the boundary, and the semiclassical backreaction affects whether the system will undergo a first-order phase transition or not. The second stage is based on $\nu_c$ as the dividing line, and at this point, the semiclassical backreaction affects the continuity of the first-order phase transition of the system itself, causing the first-order phase transition of the black hole to show a phenomenon of from existence to disappearance, then to emergence, and finally to disappearance. Visually, we may refer to it as the thermodynamic bounce effect.

The reentrant phase transition discovered in~\cite{Frassino2024a} is described by the combination of the first- and zeroth-order phase transitions as the temperature monotonically varies. To be precise, it is reentrant Hawking-Page phase transition. As the temperature monotonically increases, there are reentrant phase transitions from thermal AdS state to qBTZ black hole state and back to thermal AdS state. The former transition, since there is a discontinuity in the slope of the free energy, is a first-order phase transition, a quantum analog of the Hawking-Page phase transition. The latter transition between the qBTZ state and thermal AdS state, since there is a jump discontinuity in the free energy, is a zeroth-order phase transition. For the bounce effect in this paper, it is a phenomenon where as the temperature increases, the system undergoes a first-order phase transition, which suddenly disappears at a certain temperature and as the temperature increases again, the first-order phase transition begins again. This result is reflected in the phase transition rate, showing a ``bounce'' phenomenon. As the temperature continues to increase, the phase transition rate starts to decrease from a certain value to zero, and then starts to increase from zero to the maximum value before gradually decreasing again to zero. What we are referring to here are the states of the qBTZ black holes, similar to gas-liquid phase transitions, which can be tentatively labeled as small qBTZ black hole state and large qBTZ black hole state.

Both the reentrant phase transitions and the bounce effect do not occur for the classical BTZ black hole. For qBTZ black hole, these phenomena only occur for large $\nu$, i.e., large backreaction, for which the brane has decreasing tension and the gravitational theory on the brane becomes more massive and effectively four dimensional. The brane geometry is asymptotically AdS. The brane gravity theory has a holographic interpretation in terms of a (defect) CFT, and the phase transitions of the qBTZ black hole should have a dual interpretation, like the higher-dimensional origin of extended black hole thermodynamics through holographic braneworlds. Moreover, for the duality between the bulk and brane system, the qBTZ free energy is equal to the bulk black hole free energy, thus having the same phase behavior, the dual CFT should exhibit reentrant phase transitions and bounce effect. We think it would be interesting to study this further.

Based on our explanation of the thermodynamic bounce effect, it originates from the non-monotonicity of entropy in black hole thermodynamic systems. We believe that the results presented in the paper have a certain degree of universality. For the quantum charged black holes~\cite{Climent2024,Feng2024} and quantum dS black holes~\cite{Emparan2022,Panella2023}, it can be seen that the thermodynamic entropy of these black holes is indeed non-monotonic, so this thermodynamic bounce effect will be occurred, but precise parameter adjustments are needed. We believe that these tasks can be left for further discussion, but the results are predictable, according to our explanation of the current outcome. For the massive gravity and the black hole in higher curvature gravity, as far as we know, the thermodynamic entropy of this black hole is still a monotonic function of the horizon radius, so according to our explanation of the source of the thermodynamic bounce effect, this novel thermodynamic effect should not occur. Unless we can find that the entropy of a black hole is a non-monotonic function of the event horizon radius, it is highly likely to produce some novel thermodynamic effects, just like black holes under the effects of semi-classical quantum backreaction in braneworld setting.

In addition, the observation of this interesting phenomenon also depends on the methods we adopt, namely the construction of generalized free energy and the use of the Kramer escape rate. It can be seen that this analysis scheme for the black hole thermodynamic system will be more helpful to obtain detailed descriptions of phase transitions and dynamic processes. This provides another thermodynamic reference value for a deeper understanding of the microscopic mechanisms of black holes.

\section*{Acknowledgments}
This research is supported by National Natural Science Foundation of China (Grant No. 12105222, No. 12275216, and No. 12247103) and supported by the project of Tang Scholar in Northwest University. The author would like to thank the anonymous referee for the helpful comments that improve this work greatly.

\end{document}